\documentclass[10pt, aps, superscriptaddress, nofootinbib ,floatfix, notitlepage]{revtex4-1}
\usepackage[utf8x]{inputenc}
\pdfoutput=1
\usepackage{graphicx}
\usepackage{hyperref}
\usepackage{bm}
\usepackage{slashed}
\usepackage{geometry}
\usepackage{graphicx}
\usepackage{amssymb}
\usepackage{epstopdf}
\usepackage{amsmath}
\usepackage{xcolor}
\usepackage{adjustbox}

\newcommand{\be}{\begin{equation}}
\newcommand{\ee}{\end{equation}}
\newcommand{\bea}{\begin{eqnarray}}
\newcommand{\eea}{\end{eqnarray}}

\newcommand{\LaSapienza}{Physics Department and INFN Sezione di Roma La Sapienza,\\ Piazzale Aldo Moro 5, 00185 Roma, Italy}
\newcommand{\RomatreINFN}{Istituto Nazionale di Fisica Nucleare, Sezione di Roma Tre,\\ Via della Vasca Navale 84, I-00146 Rome, Italy}
\newcommand{\Annecy}{LAPTh, Universit\'e Savoie Mont-Blanc and CNRS, F-74941 Annecy, France}

\begin{document}

\title{What we can learn  from  the angular differential rates (only) in  semileptonic $B \to D^* \ell \nu_\ell$ decays}

\author{G.\,Martinelli}\affiliation{\LaSapienza}
\author{S.\,Simula}\affiliation{\RomatreINFN}
\author{L.\,Vittorio}\affiliation{\Annecy}
 
\begin{abstract}
We present a simple approach to the study of semileptonic  $B \to D^* \ell \nu_\ell$ decays based on the angular distributions of the final state particles only. Our approach is model independent and never requires the knowledge of $\vert V_{cb}\vert$. By studying such distributions in the case of light leptons, a comparison between results from different data sets from the Belle and BelleII Collaborations and between data and Standard Model calculations is also given for several interesting quantities. 
A  good consistency is observed  between some of the experimental results and the theoretical predictions.
\end{abstract}

\maketitle

\section{Introduction}
\label{sec:intro}

In this work we discuss a straightforward approach to the analysis of semileptonic  $B \to D^* \ell \nu_\ell$ decays based on the angular distributions of the final state particles. Although several analyses which make use the angular distributions already exist in the literature\,\cite{Tanaka:2012nw, Sakaki:2013bfa, Duraisamy:2013pia, Ivanov:2016qtw, Colangelo:2018cnj, Bigi:2017njr, Jung:2018lfu, Jaiswal:2020wer, Iguro:2020cpg, Huang:2021fuc, Fedele:2023ewe, Bordone:2024weh, Bernlochner:2024xiz}, our study as well as the one in Ref.\,\cite{Bobeth:2021lya} are only based on the angular distributions and reduce the problem to the determination of few basic parameters (five in all). These parameters encode in the most general way the contributions to the differential decay rates coming from operators present in the effective Hamiltonian either in the Standard Model (SM) or from physics Beyond the Standard Model (BSM). 
The analysis is model independent and never requires the knowledge of $\vert V_{cb}\vert$. 
In this work we analyse for the first time the angular distributions of different experimental data sets. This allows a direct comparison of the results obtained from different experiments as well as with the theoretical predictions based on the hadronic form factors (FFs) obtained from available Lattice QCD (LQCD) simulations. While in some specific cases differences (within about two standard deviations) are visible, a quite good consistency is observed between some of the experimental results and the theoretical predictions of the SM using the LQCD FFs.
The present study is limited to $B \to D^* \ell \nu_\ell$ decays with light leptons in the final states, for which possible BSM contributions have been considered in the past\,\cite{Bernlochner:2014ova, Jung:2018lfu} and also more recently\,\cite{Fedele:2023ewe, Colangelo:2024mxe, Bernlochner:2024xiz}.

Using the most general structure of the four-fold differential decay rate for semileptonic $B \to D^* \ell \nu_\ell$ decays, the five basic parameters (denoted in the following as  $ \{\eta, \eta^\prime, \delta, \epsilon, \epsilon^\prime \}$) are defined in terms of experimentally measurable quantities related to different angular distributions,
which will  be the basis of our phenomenological analysis, namely
\bea
      \label{eq:thetav}
      \frac{1}{\Gamma} \frac{d\Gamma}{d\mbox{cos}\theta_v} & = &  \frac{3}{4 (1 + \eta)} \left\{ \eta + (2 - \eta) \mbox{cos}^2\theta_v \right\} ~ , ~ \\[2mm]
      \label{eq:thetal}
       \frac{1}{\Gamma} \frac{d\Gamma}{d\mbox{cos}\theta_\ell} & = &  \frac{3}{8 (1 + \eta^\prime)} \left\{ 2 + \eta^\prime -2 \delta \mbox{cos}\theta_\ell - 
                  (2 - \eta^\prime) \mbox{cos}^2\theta_\ell \right\} ~ , ~ \\[2mm]
      \label{eq:chi}
       \frac{1}{\Gamma} \frac{d\Gamma}{d\chi} & = & \frac{1}{2 \pi} \left\{ 1 - \frac{\epsilon}{1 + \eta} \mbox{cos}2\chi - 
                  \frac{\epsilon^\prime}{1 + \eta} \mbox{sin}2\chi \right\} ~ . ~
\eea 
In order to disentangle $\delta$ from $\eta^\prime$,   a separation  of the dependence of   $1/\Gamma d\Gamma / d\mbox{cos}\theta_\ell$ on the even or odd  terms  in $\mbox{cos}\theta_\ell$ is necessary.  
In literature it is common to refer to observables like the forward-backward asymmetry $A_{FB}$, the longitudinal $D^*$-polarization fraction $F_L$,  and the two transverse asymmetries $A_{1c}$ and $ A_{9c}$~\footnote{The asymmetries $ A_{1c}$ and $ A_{9c}$ correspond  to the quantities $S_3$ and $S_9 $  respectively, as defined in Ref.\,\cite{Belle:2023xgj}, multiplied by $\pi$. $ A_{9c}$ corresponds to $A^{(1)}_T$ defined in Eq.\,(37) of Ref.\,\cite{Ivanov:2016qtw}.}.  These quantities are related to the five hadronic parameters $ \{ \eta, \eta^\prime, \delta, \epsilon, \epsilon^\prime \}$  by
\bea
     \label{eq:AFB}
     A_{FB} & = & - \frac{3}{4} \frac{\delta}{1 + \eta^\prime} ~ , ~ \\[2mm]
      \label{eq:FL}
     F_L & = &\frac{1}{1 + \eta} ~ , ~ \\[2mm]
     \label{eq:A1c}
     A_{1c}& = & - \frac{\epsilon}{1 + \eta} ~ , ~\\[2mm]
     \label{eq:A9c}
     A_{9c}& =& - \frac{\epsilon^\prime}{1 + \eta} ~ , ~
\eea
 and will be used in the present analysis. 
 
 The plan of the remainder of the paper is the following: in Sec.\,\ref{sec:Ji} we  recall the most general expression of the $B \to D^*$ differential decay rate
in the momentum transfer and in the relevant angular variables.  We then derive the expressions given in Eqs.\,(\ref{eq:thetav})-(\ref{eq:A9c});  in Sec.\,\ref{sec:SM}  we express the basic parameters $ \{ \eta, \eta^\prime, \delta, \epsilon, \epsilon^\prime \}$  in terms of the helicity amplitudes computed in the SM;
in Sec.\,\ref{sec:binning} we describe our   fit of the data from different measurements, present tables and figures containing the results and discuss their  compatibility and consistency with   the SM. The final Section contains our  conclusions.

\section{The four-fold $B \to D^*$ differential decay rate and the definition of the basic parameters}
\label{sec:Ji}

\begin{figure}[htb!]
\begin{center}
\includegraphics[scale=1.25]{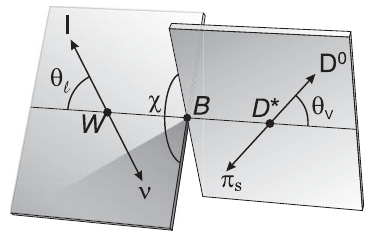}
\end{center}
\vspace{-0.5cm}
\caption{\it \small Definition of the angles $\theta_v$, $\theta_\ell$ and $\chi$ for the decay $B \to D^*(D \pi) \ell \nu_\ell$. This figure has been taken from Ref.\,\cite{Belle:2018ezy}.}
\label{fig:angles}
\end{figure}

In this section we derive the expressions in Eqs.\,(\ref{eq:thetav})-(\ref{eq:chi}) from the four-fold $B \to D^*$ differential decay rate.
The general structure of the four-fold differential rate for $B \to D^* \ell \nu_\ell$ decays, valid both within the SM  and including possible BSM effects, can be expressed in terms of twelve angular observables (coefficients) $J_i(w)$\,\cite{Duraisamy:2013pia, Bernlochner:2014ova, Ivanov:2016qtw, Belle:2023xgj}, functions of the recoil variable $w$, which is given in terms of the squared four-momentum transfer $q^2$ by
\be
    w \equiv \frac{1 + r^2 - q^2 / m_B^2}{2 r} ~ \,
    \label{eq:w}
\ee
with $r \equiv m_{D^*} / m_B$.
The dependence on the squared momentum transfer is all condensed in the angular observables themselves, which can be expressed in terms of the helicity amplitudes (and then in terms of the hadronic FFs) and of the Wilson coefficients of the relevant operators as done in Refs.\,\cite{Duraisamy:2013pia, Ivanov:2016qtw} (very detailed and complementary discussions can be also found in Refs.\,\cite{Tanaka:2012nw, Colangelo:2018cnj}). The physical quantities $J_i(w)$ are particularly relevant to scrutinize the presence of  BSM  effects in semileptonic $B \to D^*$ decays.
Following the notation of Ref.\,\cite{Belle:2023xgj} one has\footnote{The angular coefficients $J_i(w)$, defined in Eq.\,(\ref{eq:d4Gamma}), are proportional to the corresponding ones defined in Ref.\,\cite{Belle:2023xgj} by a multiplicative constant equal to $2^{10} m_B^3 / 3 m_{D^*}$.}
\bea
    \label{eq:d4Gamma}
    \frac{d^4\Gamma(B \to D^* \ell \nu_\ell)}{dw d\mbox{cos}\theta_v d\mbox{cos}\theta_\ell d\chi } & = & \frac{3}{16 \pi} \Gamma_0 \left\{ J_{1s}(w)  
        \mbox{sin}^2\theta_v + J_{1c}(w)  \mbox{cos}^2\theta_v \right. \nonumber \\[2mm]
        & + & \left.  J_{2s}(w)  \mbox{sin}^2\theta_v \mbox{cos}2\theta_\ell + J_{2c}(w) \mbox{cos}^2\theta_v \mbox{cos}2\theta_\ell \right. \nonumber \\[2mm]
        & + & \left. J_3(w) \mbox{sin}^2\theta_v \mbox{sin}^2\theta_\ell \mbox{cos}2\chi + J_4(w) \mbox{sin}2\theta_v \mbox{sin}2\theta_\ell \mbox{cos}\chi \right. \\[2mm]
        & + & \left. J_5(w) \mbox{sin}2\theta_v \mbox{sin}\theta_\ell \mbox{cos}\chi + J_{6s}(w)  \mbox{sin}^2\theta_v \mbox{cos}\theta_\ell  \right. \nonumber \\[2mm]
        & + & \left. J_{6c}(w)  \mbox{cos}^2\theta_v \mbox{cos}\theta_\ell + J_7(w) \mbox{sin}2\theta_v \mbox{sin}\theta_\ell \mbox{sin}\chi \right. \nonumber \\[2mm]
        & + & \left. J_8(w) \mbox{sin}2\theta_v \mbox{sin}2\theta_\ell \mbox{sin}\chi + J_9(w) \mbox{sin}^2\theta_v \mbox{sin}^2\theta_\ell \mbox{sin}2\chi  \right\} 
                  \nonumber ~ , ~
\eea
where
\be
    \Gamma_0 \equiv \frac{\eta_{EW}^2 m_B m_{D^*}^2}{(4 \pi)^3} G_F^2 |V_{cb}|^2 
    \label{eq:Gamma_0}
\ee
with $\eta_{EW} = 1.0066$, $G_F$ the Fermi constant and $V_{cb}$ the relevant CKM matrix element.
The angles $\theta_v$, $\theta_\ell$ and $\chi$ are defined as in Fig.~\ref{fig:angles}.

The total decay rate is given by
\be
     \label{eq:Gamma_tot}
     \Gamma(B \to D^* \ell \nu_\ell) = \frac{1}{6} \Gamma_0 \left[ 6 \overline{J}_{1s} + 3 \overline{J}_{1c} - 2 \overline{J}_{2s} - \overline{J}_{2c} \right] ~ , 
\ee
where the quantities $\overline{J}_i$ are the integrated angular coefficients over the full kinematical range of $w$, namely
\be
     \label{eq:Ji_tot}
     \overline{J}_i \equiv \int_1^{w_{max}^\ell} dw J_i(w)
\ee
with
\be
     \label{eq:wmax}
     w_{max}^\ell \equiv \frac{1 + r^2 - m_\ell ^2 / m_B^2 }{2r} ~ . ~
 \ee
By dividing Eq.\,(\ref{eq:d4Gamma}) by the total rate $\Gamma$ one gets the four-fold decay ratio independent of $V_{cb}$, namely
\bea
    \label{eq:ratio_4}
   \frac{1}{\Gamma}  \frac{d^4\Gamma(B \to D^* \ell \nu_\ell)}{dw d\mbox{cos}\theta_v d\mbox{cos}\theta_\ell d\chi } & = & \frac{1}{{\cal{N}}}  \left\{ J_{1s}(w)  
        \mbox{sin}^2\theta_v + J_{1c}(w)  \mbox{cos}^2\theta_v \right. \nonumber \\[2mm]
        & + & \left.  J_{2s}(w)  \mbox{sin}^2\theta_v \mbox{cos}2\theta_\ell + J_{2c}(w) \mbox{cos}^2\theta_v \mbox{cos}2\theta_\ell \right. \nonumber \\[2mm]
        & + & \left. J_3(w) \mbox{sin}^2\theta_v \mbox{sin}^2\theta_\ell \mbox{cos}2\chi + J_4(w) \mbox{sin}2\theta_v \mbox{sin}2\theta_\ell \mbox{cos}\chi \right. \\[2mm]
        & + & \left. J_5(w) \mbox{sin}2\theta_v \mbox{sin}\theta_\ell \mbox{cos}\chi + J_{6s}(w)  \mbox{sin}^2\theta_v \mbox{cos}\theta_\ell  \right. \nonumber \\[2mm]
        & + & \left. J_{6c}(w)  \mbox{cos}^2\theta_v \mbox{cos}\theta_\ell + J_7(w) \mbox{sin}2\theta_v \mbox{sin}\theta_\ell \mbox{sin}\chi \right. \nonumber \\[2mm]
        & + & \left. J_8(w) \mbox{sin}2\theta_v \mbox{sin}2\theta_\ell \mbox{sin}\chi + J_9(w) \mbox{sin}^2\theta_v \mbox{sin}^2\theta_\ell \mbox{sin}2\chi  \right\} 
                  \nonumber ~ , ~
\eea
where
\be
     \label{eq:norm}
     {\cal{N}} =  \frac{8 \pi}{9} \left[ 6 \overline{J}_{1s} + 3 \overline{J}_{1c} - 2 \overline{J}_{2s} - \overline{J}_{2c} \right] ~ . ~
\ee
After integrating Eq.\,(\ref{eq:ratio_4}) over the recoil variable $w$ and over two out of the three angular coordinates $\{ \theta_v, \theta_\ell, \chi \}$ we obtain the single-differential angular decay rates
\bea
      \label{eq:ratio_v_Ji}
      \frac{1}{\Gamma} \frac{d\Gamma}{d\mbox{cos}\theta_v} & = &  \frac{4 \pi}{3 {\cal{N}}} \left\{ 3 \overline{J}_{1s} - \overline{J}_{2s} \right. \nonumber \\[2mm]
              & + & \left. \left( 3 \overline{J}_{1c} - \overline{J}_{2c} - 3 \overline{J}_{1s} + \overline{J}_{2s} \right) \mbox{cos}^2\theta_v \right\} ~ , ~ \\[2mm]
      \label{eq:ratio_ell_Ji}
      \frac{1}{\Gamma} \frac{d\Gamma}{d\mbox{cos}\theta_\ell} & = & \frac{4 \pi}{3 {\cal{N}}} \left\{ 2 \overline{J}_{1s} - 2 \overline{J}_{2s} + 
                       \overline{J}_{1c} - \overline{J}_{2c} \right. \nonumber \\[2mm]
              & + & \left. \left( 2 \overline{J}_{6s} + \overline{J}_{6c} \right) \mbox{cos}\theta_\ell + 2 \left( 2 \overline{J}_{2s} + \overline{J}_{2c} \right) 
                        \mbox{cos}^2\theta_\ell \right\} ~ , ~ \\[2mm]
      \label{eq:ratio_chi_Ji}
      \frac{1}{\Gamma} \frac{d\Gamma}{d\chi} & = & \frac{1}{2 \pi} \left\{ 1 + \frac{32 \pi}{9 {\cal{N}}} \overline{J}_3 \mbox{cos}2\chi + 
                         \frac{32 \pi}{9 {\cal{N}}} \overline{J}_9 \mbox{sin}2\chi \right\} ~ . ~
\eea
By  defining  the following dimensionless quantities
\bea
     \label{eq:eta}
     \eta & \equiv & 2 \frac{3 \overline{J}_{1s} - \overline{J}_{2s}}{3 \overline{J}_{1c} - \overline{J}_{2c}} ~ , ~ \\[2mm]
      \label{eq:etap}
      \eta^\prime & \equiv & 2 \frac{2 \overline{J}_{1s} + \overline{J}_{1c} + 2 \overline{J}_{2s}  + \overline{J}_{2c}}{2 \overline{J}_{1s} + \overline{J}_{1c} - 
              3 \left( 2 \overline{J}_{2s} + \overline{J}_{2c} \right)} ~ , ~ \\[2mm]
     \label{eq:delta}
      \delta & = & - 2 \frac{2 \overline{J}_{6s} + \overline{J}_{6c}}{2 \overline{J}_{1s} + \overline{J}_{1c} - 3 \left( 2 \overline{J}_{2s} + \overline{J}_{2c} \right)} ~ , ~ \\[2mm]
     \label{eq:epsilon}
      \epsilon & \equiv & - 4 \frac{\overline{J}_3}{3 \overline{J}_{1c} - \overline{J}_{2c}} ~ , ~ \\[2mm]
     \label{eq:epsilonp}
      \epsilon^\prime & \equiv & - 4 \frac{\overline{J}_9}{3 \overline{J}_{1c} - \overline{J}_{2c}}  ~ , ~
\eea
we get  the expressions in Eqs.\,(\ref{eq:thetav})-(\ref{eq:chi}).

Therefore, even including BSM  effects (cf.~also Refs.\,\cite{Gambino:2019sif, Bobeth:2021lya}), the single-differential angular decay rates (\ref{eq:thetav})-(\ref{eq:chi}) have a precise dependence on the angular coordinates $\{ \theta_v, \theta_\ell, \chi \}$ governed only by five hadronic parameters given by $\{ \eta, \eta^\prime, \delta, \epsilon, \epsilon^\prime \}$, defined by Eqs.\,(\ref{eq:eta})-(\ref{eq:epsilonp}) in terms of the integrated angular coefficients $\overline{J}_i$.
The quantities $A_{FB}$,  $F_L$,  $A_{1c}$ and $A_{9c}$  can be easily derived from Eqs.\,(\ref{eq:thetav})-(\ref{eq:chi}) obtaining Eqs.\,(\ref {eq:AFB})-(\ref{eq:A9c}).

\section{The angular variables $J_i(w)$ in the SM}
\label{sec:SM}

Within the SM the angular coefficients $J_i(w)$ can be expressed in terms of the helicity amplitudes $H_{+, -, 0, t}(w)$\,\cite{Belle:2023xgj} as
\bea
      \label{eq:Ji_SM}
      J_{1s}(w) & = & \frac{3}{2} F(w) \left[ H_+^2(w) + H_-^2(w) \right] \left(1 + \frac{m_\ell^2}{3q^2} \right) ~ , ~ \\[2mm]
      J_{1c}(w) & = & 2 F(w) \left[ H_0^2(w) \left(1 + \frac{m_\ell^2}{q^2} \right) + 2 \frac{m_\ell^2}{q^2} H_t^2(w) \right]~ , ~ \\[2mm]
      J_{2s}(w) & = & \frac{1}{2} F(w) \left[ H_+^2(w) + H_-^2(w) \right] \left(1 - \frac{m_\ell^2}{q^2} \right) ~ , ~ \\[2mm]
      J_{2c}(w) & = & - 2 F(w) H_0^2(w) \left(1 - \frac{m_\ell^2}{q^2} \right) ~ , ~ \\[2mm]
      J_3(w) & = & - 2 F(w) H_+(w) H_-(w) \left(1 - \frac{m_\ell^2}{q^2} \right) ~ , ~ \\[2mm]
      J_4(w) & = & F(w) H_0(w)  \left[ H_+(w) + H_-(w) \right] \left(1 - \frac{m_\ell^2}{q^2} \right) ~ , ~ \\[2mm]
      J_5(w) & = & 2 F(w) \left\{ H_0(w) \left[ H_-(w) - H_+(w) \right] + \frac{m_\ell^2}{q^2} H_t(w) \left[ H_+(w) + H_-(w) \right] \right\} ~ , ~ \quad \\[2mm]
      J_{6s}(w) & = & - 2 F(w)  \left[ H_+^2(w) - H_-^2(w) \right] ~ , ~ \\[2mm]
      J_{6c}(w) & = & - 8 \frac{m_\ell^2}{q^2} F(w) H_0(w) H_t(w) ~ , ~ \\[2mm]
      J_7(w) & = & J_8(w) = J_9(w) = 0 ~ , ~
\eea
where the kinematical factor $F(w)$ is given by
\be
     \label{eq:Fw}
     F(w) \equiv \sqrt{w^2 -1} ~ \left( 1 + r^2 - 2 r w  \right) ~ \left(1 - \frac{m_\ell^2}{q^2} \right)^2 ~ . ~
\ee
It follows that, within the SM,  the quantities $\{ \eta, \eta^\prime, \delta, \epsilon, \epsilon^\prime \}$ are explicitly given by
\bea
     \label{eq:eta_SM}
     \eta & = & \frac{H_{++} + H_{--} + \frac{m_\ell^2}{2 m_B^2} \left( \widetilde{H}_{++} + \widetilde{H}_{--} \right)}
                      {H_{00} +  \frac{m_\ell^2}{2 m_B^2} \left( \widetilde{H}_{00} + 3  \widetilde{H}_{tt} \right)} ~ , ~ \\[2mm]
     \label{eq:etap_SM}
     \eta^\prime & = & \frac{H_{++} + H_{--} + \frac{m_\ell^2}{m_B^2} \left( \widetilde{H}_{00} + \widetilde{H}_{tt} \right)}
                      {H_{00} +  \frac{m_\ell^2}{2 m_B^2} \left( \widetilde{H}_{++} + \widetilde{H}_{--} -\widetilde{H}_{00} + \widetilde{H}_{tt} \right)} ~ , ~ \\[2mm]
     \label{eq:delta_SM}
     \delta & = & \frac{H_{++} - H_{--} + \frac{2m_\ell^2}{m_B^2} \widetilde{H}_{0t}}
                      {H_{00} +  \frac{m_\ell^2}{2 m_B^2} \left( \widetilde{H}_{++} + \widetilde{H}_{--} -\widetilde{H}_{00} + \widetilde{H}_{tt} \right)} ~ , ~ \\[2mm]
     \label{eq:epsilon_SM}
     \epsilon & = &  \frac{H_{+-} - \frac{m_\ell^2}{m_B^2} \widetilde{H}_{+-}}
                      {H_{00} +  \frac{m_\ell^2}{2 m_B^2} \left( \widetilde{H}_{00} + 3  \widetilde{H}_{tt} \right)} ~ , ~ \\[2mm]
     \label{eq:epsilonp_SM}
     \epsilon^\prime & = & 0 ~ , ~    
\eea
where for $i, j = \{+, -, 0, t \}$
\bea
    \label{eq:Hij}
    H_{ij} & \equiv & \int_1^{w_{max}^\ell} dw \sqrt{w^2 - 1} (1 - 2r w + r^2) ~ \left(1 - \frac{m_\ell^2}{q^2} \right)^2 ~  H_i(w) ~ H_j(w) ~ , ~ \\[2mm]
    \widetilde{H}_{ij} & \equiv & \int_1^{w_{max}^\ell} dw \sqrt{w^2 - 1} ~ \left(1 - \frac{m_\ell^2}{q^2} \right)^2 ~ H_i(w) ~ H_j(w) ~ . ~
\eea
Note that, if we neglect the mass of the final charged lepton, we have $\eta^\prime=\eta$ and the four quantities 
$A_{FB}$,  $F_L$,  $A_{1c}$ and $A_{9c}$  are sufficient to determine all the basic parameters. 

The helicity amplitudes $H_{+,-,0,t}(w)$ are related to the standard  FFs $f(w)$, $g(w)$, $F_1(w)$ and $F_2(w)$ of Ref.\,\cite{Boyd:1997kz}, corresponding to definite spin-parity (to which the unitarity bounds can be applied), by
\bea
    \label{eq:H+}
    H_+(w)  & = & f(w) - m_B^2 r \sqrt{w^2 -1} ~g(w) ~ , ~ \\[2mm]
    \label{eq:H-}
    H_-(w) & = & f(w) + m_B^2 r \sqrt{w^2 -1} ~ g(w) ~ , ~ \\[2mm]
    \label{eq:H0}
     H_0(w) & = & \frac{F_1(w)}{m_B \sqrt{1 - 2 r w + r^2}}  ~ , ~ \\[2mm]
    \label{eq:Ht}
    H_t(w) & = & \frac{m_B r \sqrt{w^2 - 1}}{\sqrt{1 - 2 r w + r^2}} ~ F_2(w) ~  . ~
\eea
In what follows, we make use of the  FFs  obtained in Ref.\,\cite{Martinelli:2023fwm} by applying the unitary Dispersive Matrix (DM) approach\,\cite{DiCarlo:2021dzg}  to  all available  LQCD results determined by FNAL/MILC\,\cite{FermilabLattice:2021cdg}, HPQCD\,\cite{Harrison:2023dzh} and JLQCD\,\cite{Aoki:2023qpa} Collaborations.   
With the above FFs we calculate the helicity amplitudes $H_{+, -, 0, t}(w)$ in the full kinematical range of $w$ (i.e., $1 \leq w \leq w_{max}^\ell$) and, consequently, the hadronic parameters $\{\eta, \eta^\prime, \delta, \epsilon\}$  through Eqs.\,(\ref{eq:eta})-(\ref{eq:epsilon}), as well as the asymmetries $A_{FB}$, $F_L$, $A_{1c}$ through Eqs.\,(\ref{eq:AFB})-(\ref{eq:A1c}). Within the SM one has $\epsilon^\prime = 0$ and, consequently, $A_{9c} = 0$.

\section{Fit of the  data and discussion of the results }
\label{sec:binning}

We now consider the three experimental data sets directly available for the single-differential decay rates $d\Gamma / dx$, where $x = \{\mbox{cos}\theta_\ell, \mbox{cos}\theta_v, \chi \}$, from Refs.\,\cite{Belle:2018ezy, Belle:2023bwv, Belle-II:2023okj}, which hereafter will be labelled as Belle18, Belle23 and BelleII23, respectively. 
For the sake of precision, only for Belle23 and BelleII23 the data sets are available directly for the ratios $(1/\Gamma) d\Gamma / dx$, while for Belle18 Ref.\,\cite{Belle:2018ezy} provides the efficiencies and response functions necessary to unfold the  measured binned yields. Using Monte Carlo samplings for propagating all the experimental uncertainties we have unfolded the data of Ref.\,\cite{Belle:2018ezy}, obtaining in this way the values of  $d\Gamma / dx$ for each experimental bin, including the corresponding covariance matrix. Our results are well consistent with those shown in Refs.\,\cite{Gambino:2019sif, Iguro:2020cpg, FermilabLattice:2021cdg, Harrison:2023dzh, Aoki:2023qpa},
For the present discussion we do not need the fourth differential decay rate $d\Gamma / dw$.

Thus, the Belle18\,\cite{Belle:2018ezy} and Belle23\,\cite{Belle:2023bwv} experimental data are given in the form of 10-bins distributions for each of the three kinematical variables $x = \{\cos \theta_l, \cos \theta_v, \chi \}$, namely
\be
      \Delta \Gamma_n^x = \int_{x_{n-1}}^{x_n} dx\prime \frac{d\Gamma}{dx\prime} ~ , \quad n = 1, 2, ... 10
      \label{eq:Gamma_bins}
\ee
with 
\bea
       \label{eq:bins}
       \{ (\mbox{cos}\theta_v)_n \} & = & \{ -1, -0.8, -0.6, -0.4,- 0.2, 0.0, 0.2, 0.4, 0.6, 0.8, 1.0\} ~ , ~ \\[2mm]
       \{ (\mbox{cos}\theta_\ell)_n \} & = & \{ -1, -0.8, -0.6, -0.4, -0.2, 0.0, 0.2, 0.4, 0.6, 0.8, 1.0\} ~ , ~\nonumber  \\[2mm]
        \{ \chi_n\} & = & \{ 0, \frac{\pi}{5}, \frac{2\pi}{5}, \frac{3\pi}{5}, \frac{4\pi}{5}, \pi, \frac{6\pi}{5}, \frac{7\pi}{5}, \frac{8\pi}{5}, \frac{9\pi}{5}, 2\pi\} ~ . ~ \nonumber
\eea
The BelleII23 data\,\cite{Belle-II:2023okj} are given in the same 10 bins for the variables $\mbox{cos}\theta_v$ and $\chi $, while in the case of $\mbox{cos}\theta_\ell$ the BelleII23 bins are only 8, since the first BelleII23 bin corresponds to the sum of the first three Belle18 and Belle23 bins and the BelleII23 bins $2 - 8$ correspond to the Belle18 and Belle23 bins $4 - 10$.
Thus, we have a total of $N = 30$ data points for both Belle18 and Belle23 and $N = 28$ data points for BelleII23, including the corresponding experimental covariance matrix of dimension $N \times N$.

For each kinematical variable $x$ the sum over the bins cover the full kinematical range. Therefore, for each set of experimental data we consider the ratios
\be
     \label{eq:ratios}
     R_n^x \equiv \frac{1}{\sum_{m = 1}^{N_x} \Delta \Gamma_m^x} \Delta \Gamma_n^x ~, ~
\ee
which should satisfy the normalization
\be
    \label{eq:norm_bins}
    \sum_{n = 1}^{N_x} R_n^x = 1\, , 
 \ee
with $N_x$ being the number of experimental bins for the variable $x$.
For the case of Belle18, using multivariate Gaussian distributions for the experimental values of $\Delta \Gamma_n^x$, we construct the ratios\,(\ref{eq:ratios}) and evaluate also the corresponding covariance matrix $\mathbf{C}_{nm}$ ($n, m = 1, ..., N$). 
Using the experimental bins\,(\ref{eq:bins}) one has (for $n = 1,2, ... 10$)
\bea
        \label{eq:ratio_v}
       R_n^{\theta_v} & = & \frac{3}{20(1+ \eta) } ~ \left[ \eta+ \frac{2 - \eta}{75} \left( 91 - 33 n + 3 n^2 \right) \right] ~ , ~ \\[2mm]
        \label{eq:ratio_ell}
        R_n^{\theta_\ell} & = & \frac{3}{40(1+ \eta^\prime)} ~ \left[ 2 + \eta^\prime - \frac{\delta}{5} \left( -11 + 2 n \right) - 
                                             \frac{2 - \eta^\prime}{75} \left( 91 - 33 n + 3 n^2 \right) \right] ~ , ~ \quad \\[2mm]
        \label{eq:ratio_chi}
        R_n^{\chi} & = & \frac{1}{10} - \frac{1}{4\pi} \frac{\epsilon}{1 + \eta} \left[ \mbox{sin}\frac{2n \pi}{5} - \mbox{sin}\frac{2(n -1) \pi}{5} \right] \nonumber \\[2mm]
                          & - &  \frac{1}{4\pi} \frac{\epsilon^\prime}{1 + \eta} \left[ \mbox{cos}\frac{2(n-1) \pi}{5} - \mbox{cos}\frac{2n \pi}{5} \right] ~ . ~
\eea

For each experiment we can now extract the values of the five hadronic parameters $\{ \eta, \eta^\prime, \delta, \epsilon, \epsilon^\prime \}$ appearing in the above equations. This is obtained by performing  a $\chi^2$-minimization procedure based on a correlated $\chi^2$. Since the covariance matrices $\mathbf{C}_{nm}$ are singular because of the conditions\,(\ref{eq:norm_bins}), we adopt the Moore-Penrose pseudoinverse approach, commonly used in least-square procedures. Since each of the matrices $\mathbf{C}_{nm}$ possesses 3 null eigenvalues, the total number of degrees of freedom is $N_{dof} = N - 3$ for each experiment.

Our results, which always correspond to the averaged $e/\mu$ case, are presented  in Table\,\ref{tab:eta} for the basic parameters $\{ \eta, \eta^\prime, \delta, \epsilon, \epsilon^\prime \}$ \,(\ref{eq:eta})-(\ref{eq:epsilonp}) and in Table\,\ref{tab:AFB} in terms of the quantities\,(\ref{eq:AFB})-(\ref{eq:A9c}). These results are given separately for the three sets of experimental data (Belle18, Belle23 and BelleII23).
In the two Tables also  other cases  have been considered, namely
\begin{itemize}
\item   Belle18 + Belle23 + BelleII23: we extract the hadronic parameters $\{ \eta, \eta^\prime, \delta, \epsilon, \epsilon^\prime \}$  from Eqs.\,(\ref{eq:ratio_v})-(\ref{eq:ratio_chi}) using simultaneously all the three experimental data sets Belle18, Belle23 and BelleII23 for the ratios (which are not correlated among different experiments);
\item Belle23(Ji) : we evaluate directly the hadronic parameters from Eqs.\,(\ref{eq:eta})-(\ref{eq:epsilonp}) using for the integrated angular coefficients $\overline{J}_i$ the sum of the experimental results in the four $w$-bins adopted in Ref.\,\cite{Belle:2023xgj}. In other words, using Eqs.\,(\ref{eq:ratio_v_Ji})-(\ref{eq:ratio_chi_Ji}) we construct a new data set for the ratios\,(\ref{eq:ratio_v})-(\ref{eq:ratio_chi}), which will be referred to as  Belle23(Ji). Note that the two sets Belle23 and Belle23(Ji) share the same four-fold differential data set. They differ only in the way the data for the single-differential angular decay rates are evaluated (see Appendix\,\ref{sec:appA});
\item LQCD :  we evaluate the SM predictions for the hadronic parameters\,(\ref{eq:eta_SM})-(\ref{eq:epsilonp_SM})  and the helicity amplitudes corresponding to the hadronic FFs obtained by the unitary DM approach in Ref.\,\cite{Martinelli:2023fwm}, based on all available LQCD results from Refs.\,\cite{FermilabLattice:2021cdg, Harrison:2023dzh, Aoki:2023qpa}\,\footnote{Very similar results can be obtained by using the unitary Boyd-Grinstein-Lebed (BGL) fit, first described in Appendix B of Ref.\,\cite{Simula:2023ujs}, performed in Ref.\,\cite{Martinelli:2023fwm} on the same LQCD data.}.
\end{itemize}

\begin{table}[htb!]
\renewcommand{\arraystretch}{1.5}
\begin{center}
\begin{adjustbox}{max width=\textwidth}
\begin{tabular}{|c||c|c|c||c|c||}
\hline
& $\eta$ & $\eta^\prime$ & $\delta$ & $\epsilon$ & $\epsilon^\prime$ \\ \hline
\hline
Belle18                                                & 0.894 (29) & 0.846 (47) & -0.534 (37) & 0.346 (28) & ~0.004 (28) \\ \hline
Belle23                                                & 1.026 (59) & 0.943 (81) & -0.595 (41) & 0.333 (61) & ~0.046 (59) \\ \hline
BelleII23                                              & 0.912 (28) & 0.908 (47) & -0.507 (28) & 0.342 (22) & ~0.005 (19) \\ \hline \hline
Belle18 + Belle23 + BelleII23              & 0.922 (18) & 0.875 (29) & -0.540 (18) & 0.337 (16) & ~0.005 (16) \\ \hline \hline
Belle23(Ji)                                           & 1.097 (73) & 0.934 (86) & -0.626 (49) & 0.361 (69) & -0.054 (67) \\ \hline \hline
LQCD                                                  & 1.109 (66) & 1.121 (66) & -0.705 (48) & 0.415 (26) & ~0.0 \\ \hline
\end{tabular}
\end{adjustbox}
\end{center}
\renewcommand{\arraystretch}{1.0}
\vspace{-0.25cm}
\caption{\it \small Results obtained for the five hadronic parameters $\{ \eta, \eta^\prime, \delta, \epsilon, \epsilon^\prime \}$, describing the dependence of the ratios \,(\ref{eq:ratio_v})-(\ref{eq:ratio_chi}) on the experimental bins of the Belle18\,\cite{Belle:2018ezy},  Belle23\,\cite{Belle:2023bwv} and BelleII23\,\cite{Belle-II:2023okj} data sets. The  row denoted by Belle18+Belle23+BelleII23 corresponds to the results obtained using simultaneously all the three experimental data sets. The  row denoted as Belle23(Ji)  shows the results corresponding to Eqs.\,(\ref{eq:eta})-(\ref{eq:epsilonp}) using the experimental results for the $w$-integrated angular coefficients $\overline{J}_i$ from Ref.\,\cite{Belle:2023xgj}. The last row shows the SM predictions\,(\ref{eq:eta_SM})-(\ref{eq:epsilonp_SM}) obtained by using  the hadronic FFs of the unitary DM approach of Ref.\,\cite{Martinelli:2023fwm} based on all available LQCD results from FNAL/MILC\,\cite{FermilabLattice:2021cdg}, HPQCD\,\cite{Harrison:2023dzh} and JLQCD\,\cite{Aoki:2023qpa} Collaborations. All the results correspond to the averaged $e/\mu$ case.}
\label{tab:eta}
\end{table}

\begin{table}[htb!]
\renewcommand{\arraystretch}{1.5}
\begin{center}
\begin{adjustbox}{max width=\textwidth}
\begin{tabular}{|c||c|c||c|c||}
\hline
& $A_{FB}$ & $F_L$ & $A_{1c}$ & $A_{9c}$ \\ \hline
\hline
Belle18                                                & 0.217 (13) & 0.528 ~(8) & -0.183 (15) & -0.002 (15)  \\ \hline
Belle23                                                & 0.230 (14) & 0.494 (14) & -0.165 (30) & -0.023 (29)  \\ \hline
BelleII23                                              & 0.200 (12) & 0.523 ~(8) & -0.179 (13) & -0.003 (10)  \\ \hline \hline
Belle18 + Belle23 + BelleII23              & 0.216 ~(7) & 0.520 ~(5) & -0.176 ~(9) & -0.003 ~(8) \\ \hline \hline
Belle23(Ji)                                           & 0.243 (14) & 0.477 (17) & -0.172 (32) & ~0.003 (32)  \\ \hline \hline
LQCD                                                  & 0.249 (10) & 0.475 (15) & -0.196 ~(7) & 0.0 \\ \hline
\end{tabular}
\end{adjustbox}
\end{center}
\renewcommand{\arraystretch}{1.0}
\vspace{-0.25cm}
\caption{\it \small Results for the quantities in Eqs.\,(\ref{eq:AFB})-(\ref{eq:A9c}).The description of the different rows is same as in Table\,\ref{tab:eta}.}
\label{tab:AFB}
\end{table}

The quality of the fits is acceptable. The values of the reduced $\chi^2$ variable, i.e.~$\chi^2 / N_{d.o.f.}$, turn out to be $\simeq 0.3$ (Belle18), $\simeq 1.2$ (Belle23), $\simeq 1.7$ (BelleII23) and $\simeq 1.0$ (Belle18 + Belle23 + BelleII23). 
The following comments are in order:
\begin{itemize}
\item the hadronic parameters and the asymmetries extracted from the Belle18 and BelleII23 data sets are  consistent within one standard deviations and more precise than those determined from the Belle23 data set. Differences not exceeding two standard deviations are visible with respect to the Belle23 results;
\item the results obtained using simultaneously all the three experimental data sets are dominated by the Belle18 and BelleII23 data sets;
\item the hadronic parameters and the asymmetries determined using the integrated angular coefficients $\overline{J}_i$ of Ref.\,\cite{Belle:2023xgj} turn out to be consistent with those corresponding to the Belle23 data set within less than one standard deviation. Such small deviations goes in the direction of increasing the differences with respect to the results obtained from the Belle18 and BelleII23 data sets;
\item  the SM predictions based on the hadronic FFs obtained by the unitary DM approach\,\cite{Martinelli:2023fwm} starting from available LQCD results, are largely consistent with the results of the Belle23 and Belle23(Ji) data sets (which are not independent), whereas they show some tensions with Belle18 and BelleII23 as well as with the average Belle18 + Belle23 + BelleII23, made over all the three experiments, except for the case of the asymmetry $A_{1c}$.
\end{itemize}

\begin{figure}[htb!]
\begin{center}
\includegraphics[scale=0.225]{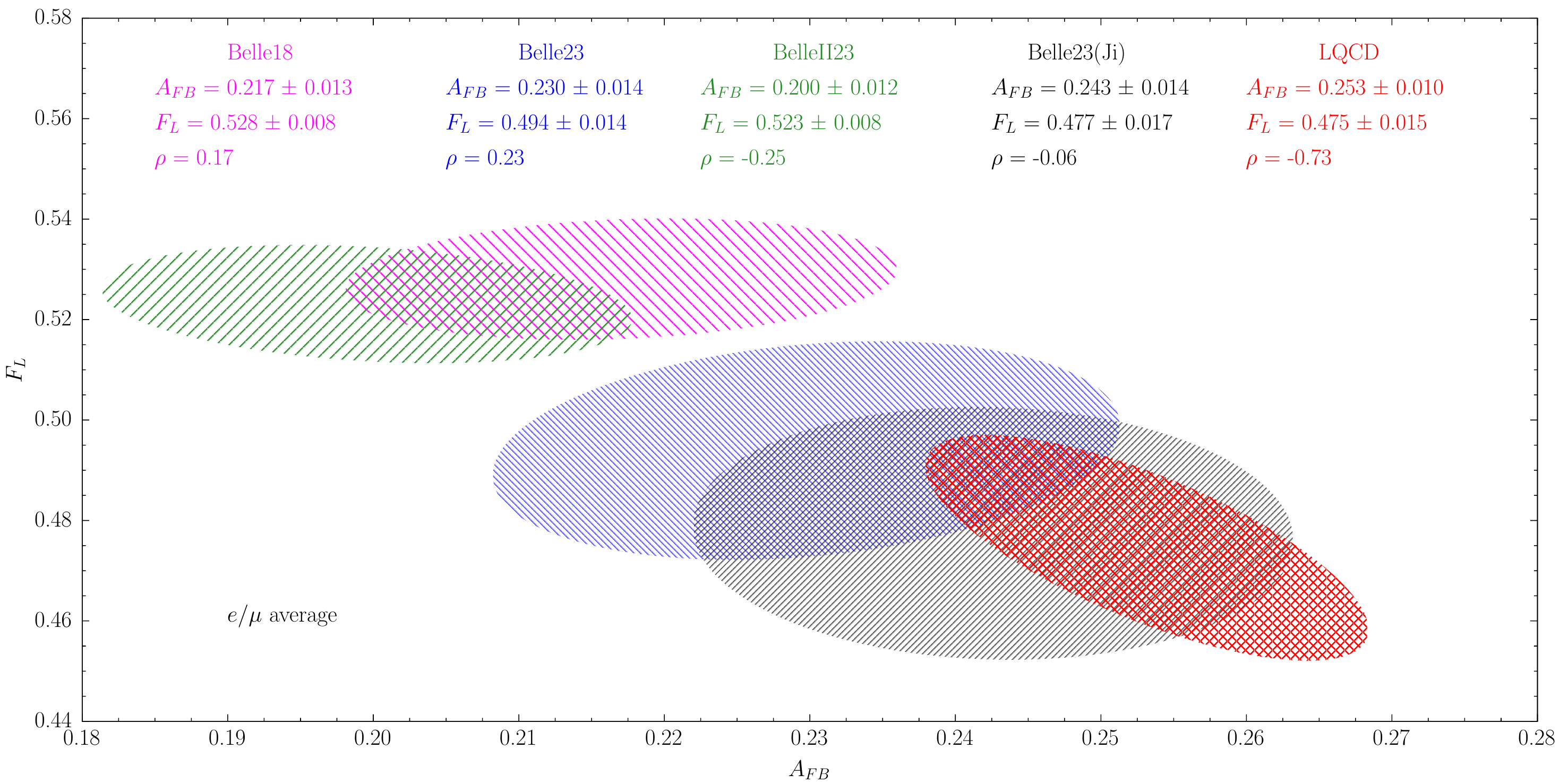}
\includegraphics[scale=0.225]{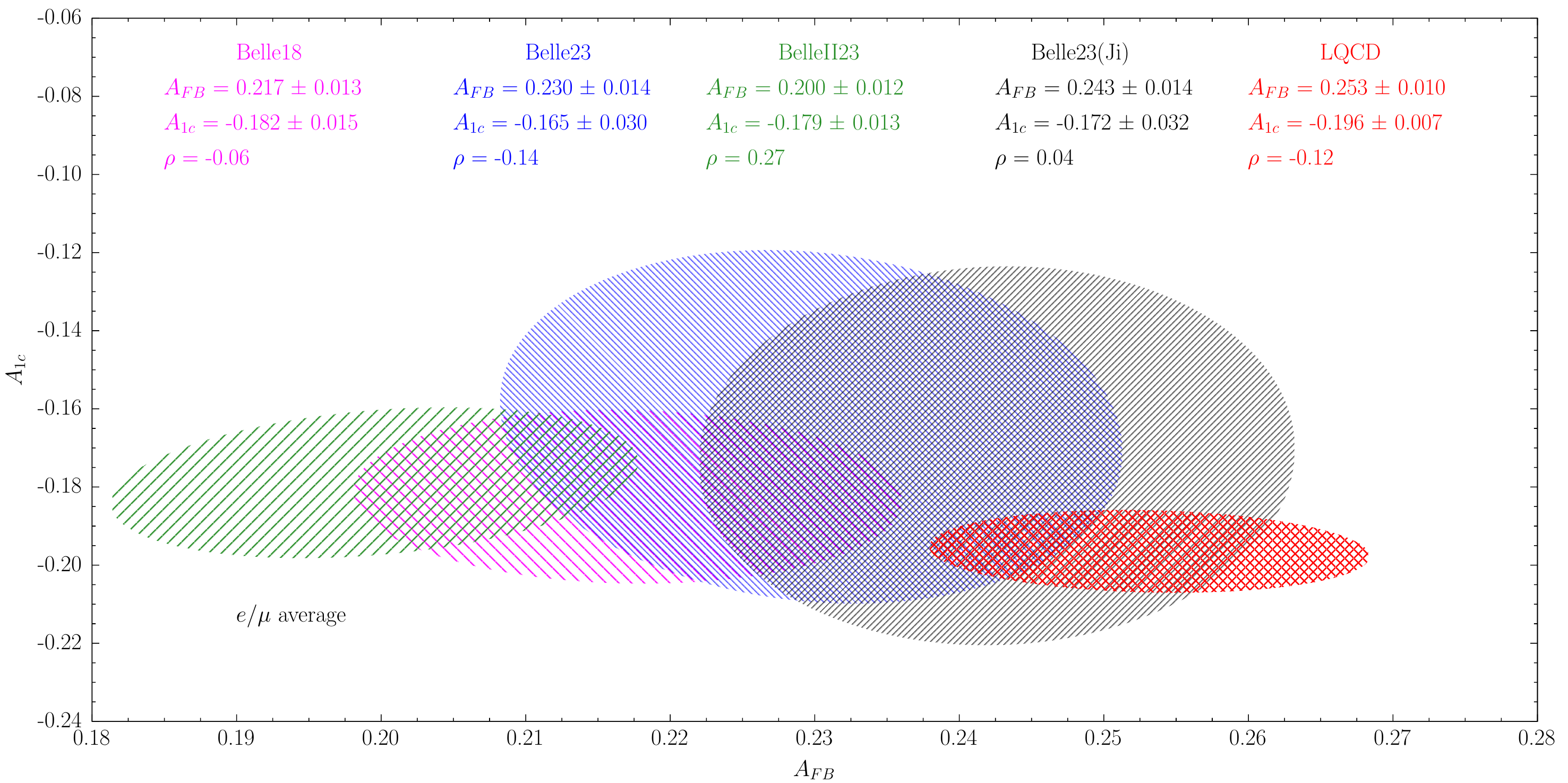}
\includegraphics[scale=0.225]{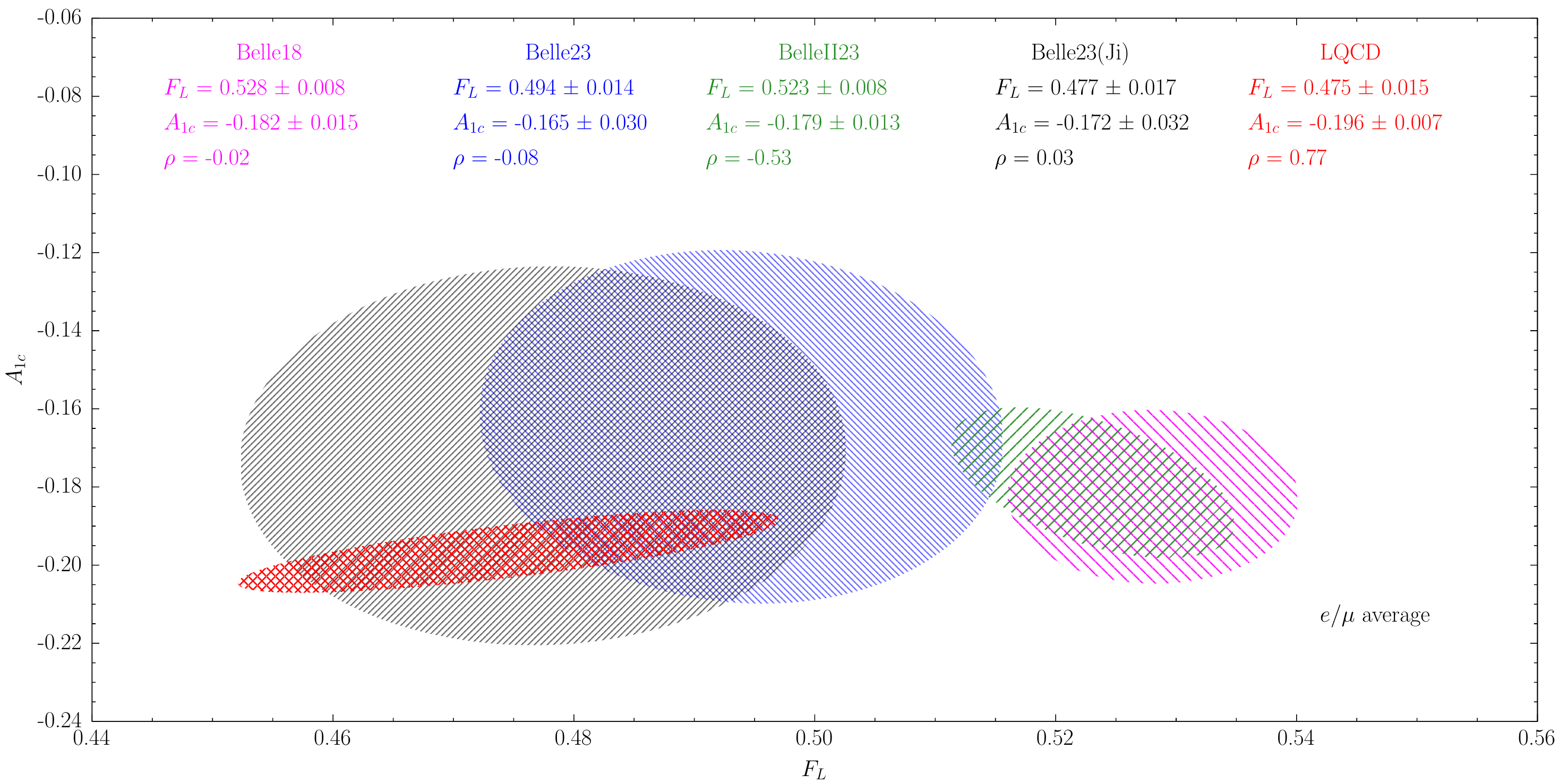}
\end{center}
\vspace{-0.5cm}
\caption{\it \small Contour plots (at $68 \%$ probability) for the asymmetries $A_{FB}$, $F_L$ and $A_{1c}$ corresponding to the analyses specified in the insets and given in Table\,\ref{tab:AFB}. In the insets the quantity $\rho$ represents the correlation coefficient.}
\label{fig:contours_light}
\end{figure}

\noindent A visual representation of the above findings, and in particular of the spread between the experimental and theoretical results, is presented in Fig.\,\ref{fig:contours_light}, where the quantities $A_{FB}$, $F_L$ and $A_{1c}$ are shown as contour plots that include the correlations among the various quantities.

\begin{figure}[htb!]
\begin{center}
\includegraphics[scale=0.35]{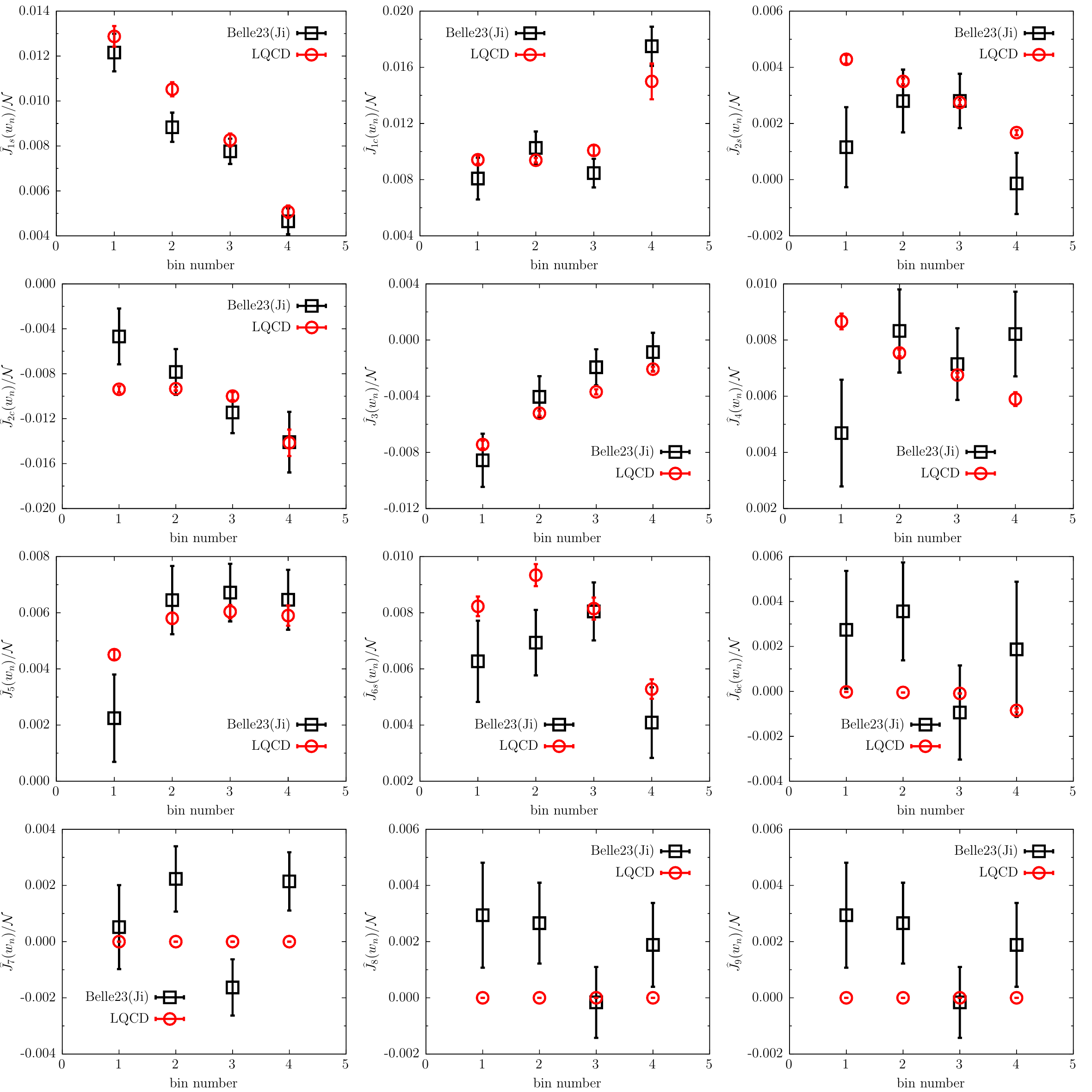}
\caption{\it \small Normalized  angular coefficients $\widehat{J}_i(w_n) /{\cal{N}}$, where ${\cal{N}}$ is given in Eq.\,(\ref{eq:norm}). The red circles represent the SM predictions corresponding to the hadronic FFs of the unitary DM approach of Ref.\,\cite{Martinelli:2023fwm} based on all available LQCD results from Refs.\,\cite{FermilabLattice:2021cdg, Harrison:2023dzh, Aoki:2023qpa}. The black squares are the experimental determinations of these quantities as measured by the Belle Collaboration in Ref.\,\cite{Belle:2023xgj}. The quantities $\widehat{J}_7(w_n), \widehat{J}_8(w_n), \widehat{J}_9(w_n)$ are exactly zero within the SM.}
\label{Jiplot}
\end{center}
\end{figure}

In Ref.\,\cite{Belle:2023xgj} the partially-integrated angular coefficients $\widehat{J}_i(w_n)$ have been determined in four $w$-bins, namely
\be
      \widehat{J}_i(w_n) \equiv \int_{w_{n-1}}^{w_n} dw J_i(w) ~ , \quad n = 1, 2, 3, 4
      \label{eq:wbins}
\ee
where $\{ w_n \} = \{1.0, 1.15, 1.25, 1.35, w_{max}^\ell \}$.
We may compare the experimental values of  $\widehat{J}_i(w_n) $ from Ref.\,\cite{Belle:2023xgj} with the the theoretical predictions  obtained  from the LQCD FFs from Ref.\,\cite{Martinelli:2023fwm}. The differences between  the theory and the experiment, which can be considered only in the case of the Belle23(Ji) data, never exceed a $2\sigma$  level. The results are presented in Fig.\,\ref{Jiplot}.

\begin{figure}[thb!]
\begin{center}
\includegraphics[scale=0.35]{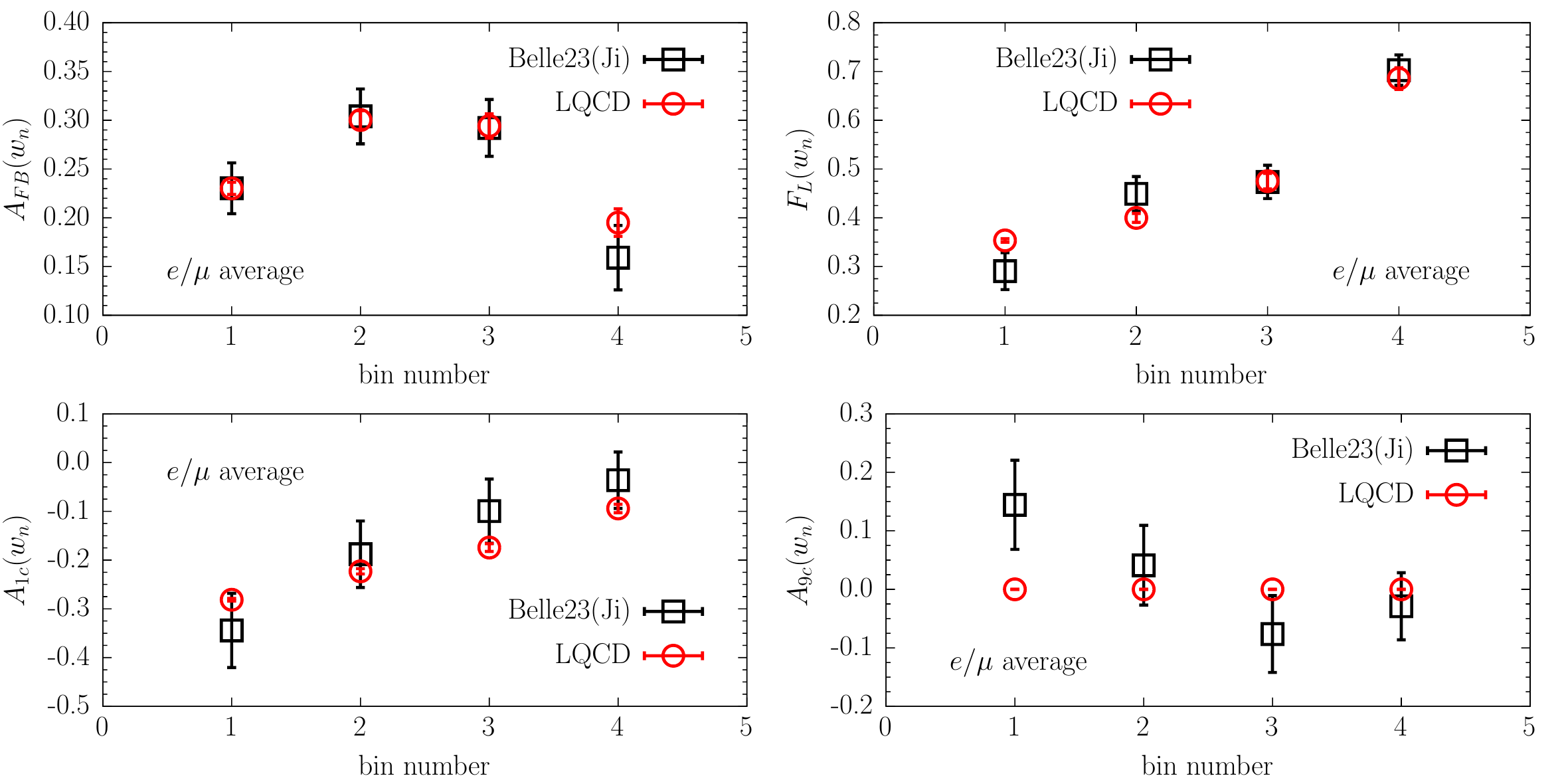}
\end{center}
\vspace{-0.5cm}
\caption{\it \small The bin-asymmetries $A_{FB}(w_n)$, $F_L(w_n)$, $A_{1c}(w_n)$ and $A_{9c}(w_n)$, evaluated separately in the four $w$-bins of Ref.\,\cite{Belle:2023xgj} (see Table\,\ref{tab:AFB_wbins}).}
\label{fig:wbins}
\end{figure}

After replacing in Eqs.\,(\ref{eq:eta})-(\ref{eq:epsilonp}) the quantities $\overline{J}_i$ with the corresponding partially-integrated ones $\widehat{J}_i(w_n)$,  the five hadronic parameters $\{ \eta, \eta^\prime, \delta, \epsilon, \epsilon^\prime \}(w_n)$ can be determined separately in each of the four $w$-bins of Ref.\,\cite{Belle:2023xgj}, as well as also the bin-quantities $A_{FB}(w_n)$, $F_L(w_n)$, $A_{1c}(w_n)$ and $A_{9c}(w_n)$, corresponding to Eqs.\,(\ref{eq:AFB})-(\ref{eq:A9c}). 
The results for $A_{FB}(w_n)$, $F_L(w_n)$, $A_{1c}(w_n)$ and $A_{9c}(w_n)$ are collected in Table\,\ref{tab:AFB_wbins} and shown in Fig.\,\ref{fig:wbins}. The experimental results  from Ref.\,\cite{Belle:2023xgj} turn out to be well consistent with the SM predictions corresponding to the hadronic FFs of the unitary DM approach of Ref.\,\cite{Martinelli:2023fwm}, based on all available LQCD results from Refs.\,\cite{FermilabLattice:2021cdg, Harrison:2023dzh, Aoki:2023qpa}.
\begin{table}[htb!]
\renewcommand{\arraystretch}{1.5}
\begin{center}
\begin{adjustbox}{max width=\textwidth}
\begin{tabular}{|c||c|c|c|c||c|c|c|c||}
\hline
w-bin & \multicolumn{4}{|c||}{Belle23(Ji)} & \multicolumn{4}{|c||}{LQCD} \\ \cline{2-9}
          & $A_{FB}$ & $F_L$ & $A_{1c}$ & $A_{9c}$  & $A_{FB}$ & $F_L$ & $A_{1c}$ & $A_{9c}$ \\ \hline \hline
1.00 - 1.15  & 0.230 (26) & 0.291 (38) & -0.344 (76) & +0.144 (76) & 0.230 (63) & 0.354 ~(4) & -0.281 (3) & -- \\ \hline
1.15 - 1.25  & 0.304 (28) & 0.449 (35) & -0.188 (68) & +0.041 (68) & 0.300 (10) & 0.400 ~(9) & -0.223 (5) & -- \\ \hline
1.25 - 1.35  & 0.292 (29) & 0.474 (34) & -0.100 (66) & -0.076 (66) & 0.292 (13) & 0.475 (16) & -0.174 (8) & -- \\ \hline
1.35 - 1.50  & 0.159 (33) & 0.703 (31) & -0.036 (58) & -0.029 (57) & 0.180 (15) & 0.686 (22) & -0.094 (9) & -- \\ \hline \hline
\end{tabular}
\end{adjustbox}
\end{center}
\renewcommand{\arraystretch}{1.0}
\vspace{-0.25cm}
\caption{\it \small The bin-asymmetries $A_{FB}(w_n)$, $F_L(w_n)$, $A_{1c}(w_n)$ and $A_{9c}(w_n)$, evaluated separately in the four $w$-bins of Ref.\,\cite{Belle:2023xgj}. Columns 2-5 correspond to the experimental results from Ref.\,\cite{Belle:2023xgj}, while columns 6-9 refer to the SM predictions corresponding to the hadronic FFs of the unitary DM approach of Ref.\,\cite{Martinelli:2023fwm} based on all available LQCD results from Refs.\,\cite{FermilabLattice:2021cdg, Harrison:2023dzh, Aoki:2023qpa}. Within the SM the asymmetry $A_{9c}(w_n)$ is exactly zero.} 
\label{tab:AFB_wbins}
\end{table}
The above findings seem to indicate that the w-dependence of the experimental angular coefficients $\widehat{J}_i(w_n)$ from Ref.\,\cite{Belle:2023xgj} is compatible, within $\simeq 2 \sigma$, with the slope of the hadronic FFs obtained in Ref.\,\cite{Martinelli:2023fwm} using the available LQCD determinations.

\section{Conclusions}
\label{sec:conclusions}

We have presented the results of an analysis of semileptonic  $B \to D^* \ell \nu_\ell$ decays based only on the angular distributions of the final leptons. In this way, the problem is reduced to the determination of five basic parameters, which encode in the most general way the contributions to the differential decay rates coming from operators present in the effective Hamiltonian either in the SM or from BSM physics. 
The analysis is model independent and never requires the knowledge of $\vert V_{cb}\vert$. 

We have analyzed for the first time the angular distributions of the experimental data sets from Refs.\,\cite{Belle:2018ezy, Belle:2023bwv, Belle-II:2023okj}. This has allowed a direct comparison of the results obtained from different experiments as well as with the theoretical predictions based on the hadronic FFs obtained from LQCD simulations. 
We have shown that for $A_{FB}$, $F_L$ and $A_{1c}$ there are visible differences between different experimental data sets within about two standard deviations {\bf (see Fig.\,\ref{fig:contours_light}).} Similar differences exist between the SM predictions, based on all the FFs available from LQCD, and some sets of data. 

A remarkable good agreement is observed between the experimental data of Ref.\,\cite{Belle:2023xgj}, given in terms of the angular coefficients $J_i(w)$ (see Eq.\,(\ref{eq:d4Gamma})), and the SM LQCD predictions, as shown in Figs.\,\ref{Jiplot} and \ref{fig:wbins}. Such a consistency, once confirmed by further experiments, may leave little room to BSM effects in the light lepton sector. In this respect, the forthcoming results from LHCb\,\cite{LHCb} concerning the coefficients $J_i(w)$ in the case of $B_s^0 \to D_s^* \mu \nu_\mu$ decays will be very valuable.

Finally, we mention that in this work our approach has been applied to the case of the decay data for final light leptons. It can be clearly extended to the case of final $\tau$ leptons once experimental data will be available.

\section*{Acknowledgements}
The authors warmly thanks M.\,Valli for his valuable help in performing the unfolding of the data of Ref.\,\cite{Belle:2018ezy}.
S.S.~is supported by the Italian Ministry of Research (MIUR) under grant PRIN 2022N4W8WR. The work of L.V.~is supported by the French Agence Nationale de la Recherche (ANR) under contracts ANR-19-CE31-0016 (‘GammaRare’) and ANR-23-CE31-0018 (‘InvISYble’).

\appendix

\section{The Belle23 and Belle23(Ji) data sets}
\label{sec:appA}

The two data sets Belle23 and Belle23(Ji) share the same raw data for the four-fold differential decay rate, i.e.~the l.h.s.~of Eq.\,(\ref{eq:d4Gamma}).
However, they differ in the way the ratios $(1 / \Gamma) d\Gamma / dx$ with $x = \{w, \mbox{cos}\theta_\ell, \mbox{cos}\theta_v, \chi \}$ are evaluated.

In the case of Belle23 the ratios have been obtained directly in Ref.\,\cite{Belle:2023bwv} by integrating the raw data of the four-fold differential decay rate over three out of the four kinematical variables $x$.

In the case of Belle23(Ji) two steps are involved. The first one has been performed directly in Ref.\,\cite{Belle:2023xgj} and it is the extraction of the angular coefficients $J_i$ obtained by fitting the raw data of the four-fold differential decay rate through Eq.\,(\ref{eq:d4Gamma}) using four $w$-bins.  
The second step has been performed in this work and it is the evaluation of the ratios (\ref{eq:ratio_v_Ji})-(\ref{eq:ratio_chi_Ji}) using the integrated angular coefficients $\overline{J}_i$, as described in Section\,\ref{sec:binning}.

\begin{figure}[htb!]
\begin{center}
\includegraphics[scale=0.325]{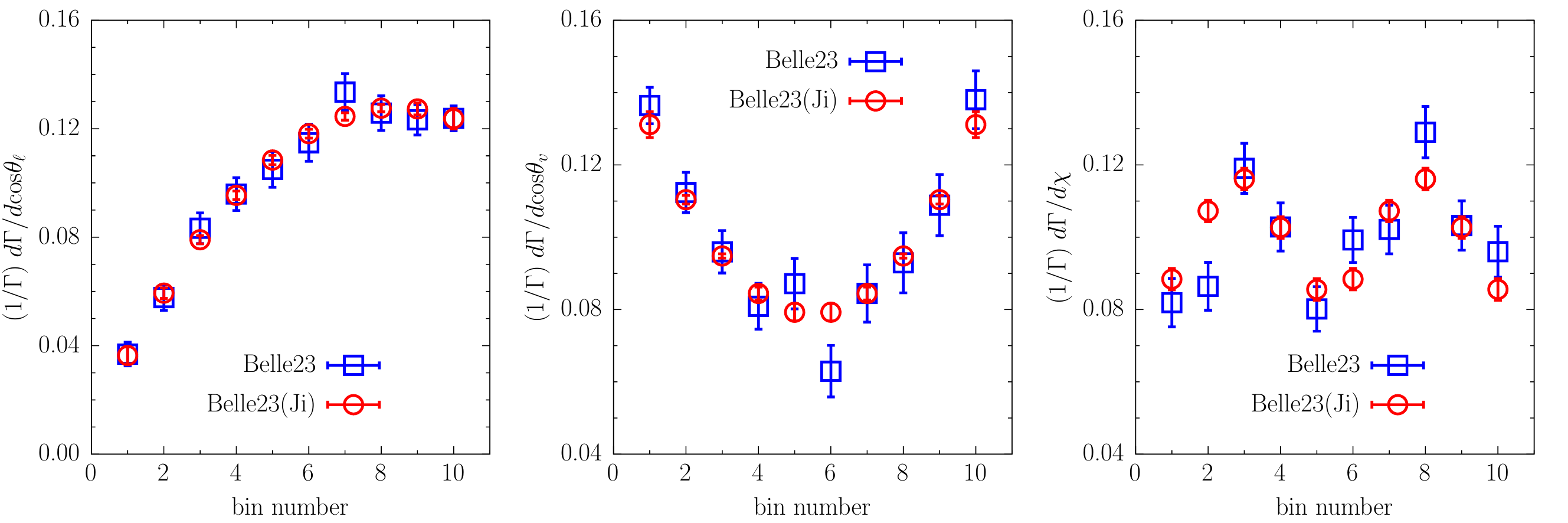}
\end{center}
\vspace{-0.5cm}
\caption{\it \small Ratios $(1 / \Gamma) d\Gamma / dx$ for $x = \{\mbox{cos}\theta_\ell, \mbox{cos}\theta_v, \chi \}$ corresponding to the two data sets Belle23 and Belle23(Ji). The Belle23 data are directly available from Ref.\,\cite{Belle:2023bwv}, while the Belle23(Ji) data have been evaluated in this work using in Eqs.\,(\ref{eq:ratio_v_Ji})-(\ref{eq:ratio_chi_Ji}) the integrated angular coefficients $\overline{J}_i$, given by Eq.\,(\ref{eq:Ji_tot}), corresponding to the results of Ref.\,\cite{Belle:2023xgj}.}
\label{fig:comparison}
\end{figure}

Note that the uncertainties and correlations of the raw data of the four-fold differential decay rate may have a different impact in the two procedures corresponding to Belle23 and Belle23(Ji).
This is confirmed in Fig.\,\ref{fig:comparison}, where the results for the angular ratios are explicitly shown. 
The two data sets, Belle23 and Belle23(Ji), are not equivalent for the ratios and, therefore, the values of the extracted hadronic parameters as well as those of $A_{FB}$, $F_L$, $A_{1c}$ and $A_{9c}$ may differ.
We stress, however, that, as shown in Fig.\,\ref{fig:contours_light}, the differences in the hadronic parameters corresponding to Belle23 and Belle23(Ji) are well below one standard deviation.

\bibliography{biblio}
\bibliographystyle{JHEP}

\end{document}